\documentclass[12pt,a4paper]{article}
\usepackage{amssymb}
\usepackage[dvips]{epsfig}
\def\beq{\begin{equation}}
\def\eeq{\end{equation}}
\def\beqr{\begin{eqnarray}}
\def\eeqr{\end{eqnarray}}
\def\nuc#1#2#3 {Nucl. Phys. {\bf#1}, #2 (#3)}
\def\mpla#1#2#3 {Mod. Phys. Lett. A {\bf#1}, #2 (#3)}
\def\plb#1#2#3 {Phys. Lett. B {\bf#1}, #2 (#3)}
\def\prd#1#2#3 {Phys. Rev. D {\bf#1}, #2 (#3)}
\def\prl#1#2#3 {Phys. Rev. Lett. {\bf#1}, #2 (#3)}
\def\ptp#1#2#3 {Prog. Theor. Phys. {\bf#1}, #2 (#3)}
\def\zpc#1#2#3 {Z. Phys. C {\bf#1}, #2 (#3)}
\def\ibid#1#2#3 {{\it ibid.} {\bf#1}, #2 (#3)}
\def\none#1#2#3 {{\bf#1}, #2 (#3)}
\begin{document}
\begin{titlepage}
\begin{flushright}
TPI-MINN-02/16 \\
June 2002
\end{flushright}
\vspace{0.7in}
\begin{center}
{\Large \bf $B\bar{B}$ Mixing and $CP$ Violation
  in $SU(2)_L \times SU(2)_R \times U(1)$ Models \\ }
\vspace{1.2in}
{\bf Soo-hyeon Nam \\ }
Theoretical Physics Institute, University of Minnesota \\
Minneapolis, MN 55455 \\
\vspace{1.2in}
{\bf Abstract \\ }
\end{center}
We reexamine the mass mixing and $CP$ violation in the $B\bar{B}$ system in general
$SU(2)_L \times SU(2)_R \times U(1)$ models related to the recent measurements.
The right-handed contributions can be sizable in $B\bar{B}$ mixing and $CP$ asymmetry
in $B$ decays for a heavy $W^\prime$ even with a mass about 3 TeV.
On the other hand the lower bound on the mass of $W^\prime$ can be taken down to
approximately 300 GeV.
\end{titlepage}

\section{Introduction}

\begin{figure}[ht]
 \vskip3mm
 \epsfxsize=7cm
 \centerline{\epsfbox{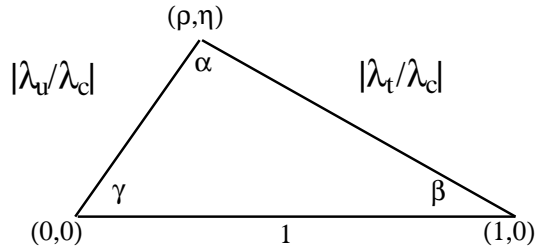}}
 \caption{Unitary triangle ($\lambda_i = U_{id}^\ast U_{ib}$).}
 \label{fig:triangle}
\end{figure}

  The Standard $SU(2)_L\times U(1)$ model ($SM$) has been very successful in describing
the known weak interaction phenomena.  But the consistency of the present experimental results
with the general scheme of charged weak interactions and $CP$ violation in the $SM$ is non-trivial
so the model is challenged both experimentally and theoretically in its prediction of
large $CP$ violation effects in the B meson system \cite{Review}.
As one of the simplest extensions of the Standard model gauge group
and so the complement of the purely left-handed nature of the $SM$, the left-right theory
with the group $SU(2)_L \times SU(2)_R \times U(1)$ has been widely studied.
In this model, even with two generations of quarks one could get $CP$ violation. With
three generations of quarks, this model contains many parameters and many sources of
$CP$ violation \cite{Pati74}.  One of the main source is the relative phase $\alpha$
between the two vacuum expectation values (VEVs) $k$ and $k^\prime$ of
the Higgs bidoublet $\Phi$.  The other sources are the complex phases in the left
and right-handed quark mixing matrices $U^L$ and $U^R$ respectively.  Here it would be
convenient to regard $U^L$ as the usual Cabbibo-Kobayashi-Maskawa ($CKM$) matrix
and shift all phases except one to $U^R$.  Using the Wolfenstein parameterization
\cite{Wolfen}, we can express the $CKM$ matrix approximately as
\beq
U^L = \left( \begin{array}{ccc} 1-\lambda^2/2 & \lambda & A\lambda^3(\rho-i\eta) \\
  -\lambda & 1-\lambda^2/2 & A\lambda^2 \\ A\lambda^3(1-\rho-i\eta) & -A\lambda^2 & 1
  \end{array} \right) \ + \ O(\lambda^4),
\eeq
where $\lambda\ (\approx 0.22)$ is an real expansion parameter, and $A$, $\rho$ and $\eta$
are also real quantities.  From the above expression, the elements $U^L_{ub}$ and $U^L_{td}$
can be parameterized in terms of two phases $\gamma$ and $\beta$ respectively which form
a unitary triangle (Fig.\ref{fig:triangle}) given by the orthogonality condition
$\sum_{i=u,c,t}U_{id}U_{ib}^* = 0$.
The recent significant measurements of $\beta_{exp}$ give \cite{Experiment}
\beq \label{beta-exp}
\sin 2\beta_{exp} = \left\{ \begin{array}{ll}
  0.59 \pm 0.14 \pm 0.05 & \mbox{\ \ ({\sl BABAR})}\\
  0.99 \pm 0.14 \pm 0.06 & \mbox{\ \ (Belle)} \end{array} \right. .
\eeq
If there are new physics effects involved, the experimental value $\beta_{exp}$
can be expressed through other parameters representing
the new physics as well as the phase of $U^L_{td}$ in the $SM$.

 In addition to the phases mentioned above, the masses ($M_{W_R}$) of the right-handed gauge bosons,
the mixing angle $\xi$ between the left- and right-handed gauge bosons $W_L$ and $W_R$,
and the right-handed gauge coupling constant $g_R$ play an important role
in new physics effects as fundamental input parameters in the left-right model ($LRM$).
The success of the $SM$ in the low-energy phenomenology requires that the masses ($M_{W_R}$)
of the right-handed gauge bosons are significantly larger than those ($M_{W_L}$) of
left-handed gauge bosons. The first lower bound on $M_{W_R}$ comes from a study of
the low energy charged current sector allowing $M_{W_R}\gtrsim 3M_{W_L}\approx 240$ GeV
\cite{Beg77}.  Soon after, many theoretical limits have been presented on $M_{W_R}$
and $\xi$ under various assumptions \cite{Lang}.  The recent experimental limits are
obtained by D\O\ and CDF from the direct search of the decay channels of the extra
gauge bosons $W^{\prime +} \rightarrow \ell^+_R \nu_R$.  D\O\ found
$M_{W^\prime} > 720$ GeV for $m_{\nu_R} \ll M_{W^\prime}$ or
$M_{W^\prime} > 650$ GeV for $m_{\nu_R} = M_{W^\prime}/2$ \cite{D0}.
CDF has a limit of $M_{W^\prime} > 652$ GeV for $m_{\nu_R} \ll M_{W^\prime}$ if $\nu_R$
is stable \cite{CDF}.  All of these limits were obtained assuming manifest ($U^R=U^L$) or
pseudo-manifest ($U^R=U^{L\ast}K$) left-right symmetry ($g_L=g_R$), where $K$ is a diagonal
phase matrix \cite{Moha92}. In this paper, we will not impose the discrete left-right symmetry
which can cause troubles in explaining the cosmological baryon asymmetry and may lead to
cosmological domain-wall problems \cite{Lang2}.  However we will also consider
the possibility of the left-right symmetric case among other possibilities.

 The main purpose of this paper is to investigate the $CP$ violation in the $B^0\bar{B^0}$
system in the $LRM$ related to the recent experiments, since $B^0\bar{B^0}$ mixing has recently been
advocated as a very sensitive probe for the $CP$ violation and the presence of right-handed
current.  The $SM$ contribution to $K^0\bar{K^0}$ mixing was previously computed
for any internal quark mass by Inami and Lim \cite{Inami}.
The right-handed contribution in the $LRM$ has been done first by
Beall, Bander, and Soni assuming the discrete left-right symmetry \cite{Beall} and again by
many authors \cite{Moha83} under various assumptions.  But we notice that the contributions
of the mixing angle $\xi$ to $B^0\bar{B^0}$ mixing and $CP$ asymmetry can be large
due to the heaviness of the top quark mass and the possibility of the enhancement
in the right-handed quark mixing matrix in the general $LRM$.  After reviewing the structure of
the $LRM$ in Sec.2, we will discuss $B^0\bar{B^0}$ mixing in Sec.3 and
$CP$ asymmetry in $B^0$ decay in Sec.4 in detail.

\section{$SU(2)_L \times SU(2)_R \times U(1)$ models}
We briefly review here some of the main features of the $LRM$, which are needed to obtain our results.
As the simplest extension of the $SM$, the gauge group of the $LRM$ breaks down to that of the $SM$ and
it finally cascades down to $U(1)_{EM}$.  The covariant derivative for the fermions $f_{L,R}$
with respect to the gauge group of the $LRM$ appears as
\beq
D^{\mu}f_{L,R} = \partial^{\mu}f_{L,R} + ig_{L,R}W_{L,R}^{\mu a} T_{L,R}^af_{L,R}
  + ig_1 B^{\mu}Sf_{L,R} .
\eeq
The electric charge which is the unbroken $U(1)$ generator is given as
\beq
Q = T_L^3 + T_R^3 + S .
\eeq
The quarks and leptons transform under the gauge group of the $LRM$ $(T_L,T_R,S)$ as
\beqr
q_L^{\prime} &=& \left( \begin{array}{c} u^{\prime} \\
  d^{\prime} \end{array} \right)_L \sim (\frac{1}{2},0,\frac{1}{6}),  \quad
q_R^{\prime} = \left( \begin{array}{c} u^{\prime} \\
  d^{\prime} \end{array} \right)_R \sim (0,\frac{1}{2},\frac{1}{6}) ,  \nonumber \\
l_L^{\prime} &=& \left( \begin{array}{c} \nu^{\prime} \\
  e^{\prime} \end{array} \right)_L \sim (\frac{1}{2},0,-\frac{1}{2}),  \quad
l_R^{\prime} = \left( \begin{array}{c} \nu^{\prime} \\
  e^{\prime} \end{array} \right)_R \sim (0,\frac{1}{2},-\frac{1}{2}),
\eeqr
where the primes indicate that the fermions are gauge rather than mass eigenstates.

In order to generate masses for the fermions and implement the symmetry breaking,
we need to include scalar fields into our theory. The simplest choice is to introduce
one Higgs multiplet and two doublets,
\beq
\Phi = \left( \begin{array}{cc} \phi_1^0 & \phi_1^+ \\ \phi_2^- & \phi_2^0 \end{array}
\right) \sim (\frac{1}{2}, \frac{1}{2}^*, 0), \
\chi_L = \left( \begin{array}{c} \chi^+ \\ \chi^0 \end{array} \right)_L
 \sim (\frac{1}{2},0,\frac{1}{2}), \
\chi_R = \left( \begin{array}{c} \chi^+ \\ \chi^0 \end{array} \right)_R
 \sim (0,\frac{1}{2},\frac{1}{2}) ,
\eeq
which acquire the vacuum expectation values (VEVs)
\beq
\langle\Phi\rangle = \left( \begin{array}{cc} k & 0 \\ 0 & k^{\prime} \end{array} \right), \quad
\langle\chi_L\rangle = \left( \begin{array}{c} 0 \\ v_L \end{array} \right), \quad
\langle\chi_R\rangle = \left( \begin{array}{c} 0 \\ v_R \end{array} \right),
\eeq
where $k$ and $k^\prime$ are complex, and $v_L$ and $v_R$ are real.
$\chi_R$ is needed to generate a large $M_{W_R}$ if $v_R \gg |k|,|k^\prime|,v_L$.
But $\chi_L$ is not essential unless we impose an left-right symmetry.  It is also possible to
adopt other choice of Higgs such as Higgs triplets instead \cite{Moha80}.
The Lagrangian for the scalar field is
\beq
L_{scalar} = Tr[(D^{\mu}\Phi)^{\dag}D_{\mu}\Phi] + (D^{\mu}\chi_L)^{\dag}D_{\mu}\chi_L
 +(D^{\mu}\chi_R)^{\dag}D_{\mu}\chi_R - V(\Phi,\chi_L,\chi_R) .
\eeq
For the Higgs fields described above, the kinetic terms in the Lagrangian generate the charged
$W$ boson matrix
\beq
M^2_{W^{\pm}} = \left( \begin{array}{cc} g^2_L(v^2_L+K^2)/2 & -g_Lg_Rk^*k^\prime \\
                       -g_Lg_Rkk^{\prime*} & g^2_R(v^2_R+K^2)/2 \end{array} \right) \equiv
                \left( \begin{array}{cc} M^2_{W_L} & M^2_{W_{LR}}e^{i\alpha} \\
                        M^2_{W_{LR}}e^{-i\alpha} & M^2_{W_R} \end{array} \right) ,
\eeq
where $K^2=|k|^2+|k^\prime|^2$ and $\alpha$ is the phase of $k^*k^\prime$.
After the mass matrix is diagonalized by a unitary transformation the eigenvalues
can be expressed in terms of a mixing angle as
\beqr
M^2_W &=& M^2_{W_L}\cos^2\xi + M^2_{W_R}\sin^2\xi + M^2_{W_{LR}}\sin2\xi , \nonumber \\
M^2_{W^\prime} &=& M^2_{W_L}\sin^2\xi + M^2_{W_R}\cos^2\xi - M^2_{W_{LR}}\sin2\xi .
\eeqr
Thus the mass eigenstates are written as
\beq
\left( \begin{array}{c} W^+ \\ W^{\prime +} \end{array} \right) =
 \left( \begin{array}{cc} \cos\xi & e^{-i\alpha}\sin\xi \\
                        -\sin\xi & e^{-i\alpha}\cos\xi \end{array} \right)
 \left( \begin{array}{c} W^+_L \\ W^+_R \end{array} \right) ,
\eeq
where $\xi$ is a mixing angle defined by
\beq
\tan 2\xi = -\frac{2M^2_{W_{LR}}}{M^2_{W_R} - M^2_{W_L}} \ .
\eeq
For $v_R \gg |k|,|k^\prime|,v_L$, the mass eigenvalues and the mixing angle reduce to
\beq
M^2_W \approx \frac{1}{2}g^2_L(v_L^2 + K^2), \quad
M^2_{W^\prime} \approx \frac{1}{2}g^2_Rv_R^2, \quad
 \xi \approx \frac{2g_L|k^*k^\prime|}{g_Rv_R^2}.
\eeq Here, the Schwarz inequality requires that $\zeta \equiv
M_W^2/M_{W^\prime}^2 \geq \xi_g \equiv (g_L/g_R)\xi$ .  From the
limits on deviations of muon decay parameters from the V-A
prediction, the lower bound on $M_{W^\prime}$ can be obtained as follows \cite{Balke88}
\beq
(g_R/g_L)^2\zeta < 0.033 \qquad \textrm{or} \qquad M_{W^\prime} > (g_R/g_L) \times 440\
 \textrm{GeV} .
\label{MWRbound}
\eeq
We will use this number for our numerical analysis.

\begin{figure}[ht]
 \vskip3mm
 \epsfxsize=10cm
 \centerline{\epsfbox{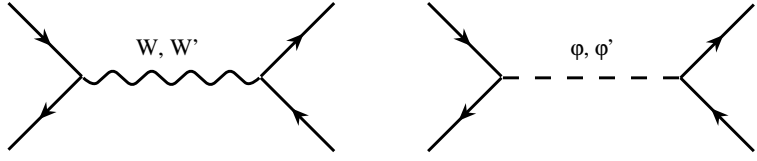}}
 \caption{Tree-level Feynman diagrams for the gauge-bosons ($W,W^\prime$)
  and the Goldstone bosons ($\varphi,\varphi^\prime$) exchange.}
 \label{fig:tree}
\end{figure}

As well as the above charged gauge bosons, the charged would-be Goldstone bosons
corresponding to the longitudinal components of the physical bosons take part in the charged
current interactions.  The coupling of the Goldstone fields to the fermions can be found
from the detailed structure of the Higgs potential $V(\Phi,\chi_{L},\chi_{R})$ and
the Yukawa couplings.  However one can directly determine the Goldstone couplings
in terms of the gauge couplings without considering the Higgs potential, but using
the Ward identities that ensure that the unphysical poles in the two diagrams
shown in Fig.\ref{fig:tree} should cancel each other \cite{Fuji72}.
The charged interaction Lagrangian is then given by
\beqr
L_{CC} &=& -\frac{1}{\sqrt{2}}\overline{P}\gamma^\mu \biggl\{[U^Lg_Lc_\xi L + U^Rg_Rs_\xi^+R]W^+_\mu
        + [-U^Lg_Ls_\xi L + U^Rg_Rc_\xi^+R]W^{\prime+}_\mu \cr
       &&+[(U^LM_Pg_Lc_\xi - U^RM_Ng_Rs_\xi^+)L
        + (-U^LM_Ng_Lc_\xi + U^RM_Pg_Rs_\xi^+)R]\frac{\varphi^+_\mu}{M_W} \\
       &&+[-(U^LM_Pg_Ls_\xi + U^RM_Ng_Rc_\xi^+)L
       + (U^LM_Ng_Ls_\xi + U^RM_Pg_Rc_\xi^+)R]\frac{\varphi^{\prime +}_\mu}{M_{W^\prime}}\biggr\}N
       \cr &&+ h.c. +  \ldots \nonumber,
\label{C-int}
\eeqr
where $c_\xi\ (s_\xi)\equiv \cos\xi\ (\sin\xi)$, $s^\pm_\xi\equiv e^{\pm\alpha}\sin\xi$,
 $L,R\equiv (1\mp\gamma^5)/2$ denote left- and right-handed projection operators,
 $M_P=diag(m_u,m_c,m_t)$ and $M_N=diag(m_d,m_s,m_b)$ are the diagonalized
 quark mass matrices, $P\ (N)$ is the mass eigenstate corresponding to its
 eigenvalue $M_P\ (M_N)$, and $U^L\ (U^R)$ is the left (right)-handed quark
 mixing matrix.

\begin{figure}[ht]
 \vskip3mm
 \epsfxsize=12cm
 \centerline{\epsfbox{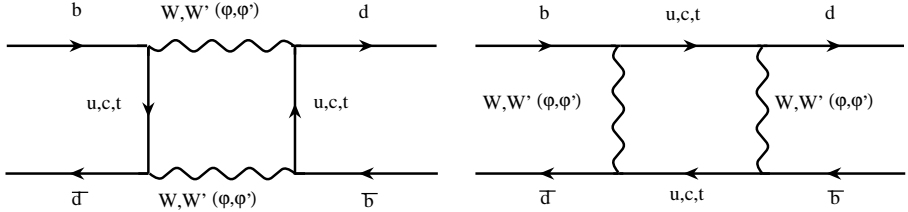}}
 \caption{Box diagrams for $B^0\bar{B^0}$ mixing with the gauge-bosons ($W,W^\prime$) and
          the Goldstone bosons ($\varphi,\varphi^\prime$).}
 \label{fig:BBbar}
\end{figure}

\section{$B^0\bar{B^0}$ mixing}
The effective Hamiltonian in the $B^0\bar{B^0}$ system is obtained by integrating out
the internal loop in the box diagrams in Fig.\ref{fig:BBbar} just as in the $SM$.
We neglect external momenta and $d$-quark mass, but the result is valid for general
internal quark masses.  One finds, using the Feynman-'t Hooft gauge,
the charged gauge boson and Goldstone boson contributions to $B^0\bar{B^0}$ mixing
in a straightforward manner :
\beq
H^{B\bar{B}}_{eff} = H^{SM}_{eff} + H^{RR}_{eff} + H^{LR}_{eff}
\eeq
with
\beqr
H^{SM}_{eff} &=& \frac{G_F^2M_W^2}{4\pi^2}\sum_{i,j=u,c,t}\lambda_i^{LL}
  \lambda_j^{LL} \biggl\{[(1 + \frac{x_i^2x_j^2}{4})f(x_i^2,x_j^2;1)
  - 2x_i^2x_j^2g(x_i^2,x_j^2;1)](\bar{d_L}\gamma_\mu b_L)^2  \cr
 &&\hspace{1.5in} +\  x_b^2x_i^2x_j^2g(x_i^2,x_j^2;1)(\bar{d_L}b_R)^2 \biggr\} , \label{Eff-HSM} \\
H^{RR}_{eff} &=& \frac{G_F^2M_W^2}{4\pi^2}(\frac{g_R}{g_L})^4 \sum_{i,j=u,c,t}
      \lambda_i^{RR} \lambda_j^{RR} \zeta f(x_i^2\zeta,x_j^2\zeta;1)(\bar{d_R}\gamma_\mu b_R)^2 , \label{Eff-HRR} \\
H^{LR}_{eff} &=& \frac{G_F^2M_W^2}{2\pi^2}(\frac{g_R}{g_L})^2 \sum_{i,j=u,c,t}
      \biggl\{ \lambda_i^{LR} \lambda_j^{RL}x_ix_j\zeta
      [4g(x_i^2,x_j^2;\zeta) - f(x_i^2,x_j^2;\zeta)](\bar{d_L}b_R)(\bar{d_R}b_L) \cr
 &&\hspace{1.1in} +\ \lambda_i^{LL} \lambda_j^{LR}x_jx_b\xi_g^{+}
      [x_i^2(g(x_i^2,x_j^2;1) - \frac{1}{4}f(x_i^2,x_j^2;1))(\bar{d_L}\gamma_\mu b_L)^2 \cr
  &&\hspace{2.2in} +\ (f(x_i^2,x_j^2;1) - x_i^2x_j^2g(x_i^2,x_j^2;1))(\bar{d_L}b_R)^2] \label{Eff-HLR} \\
 &&\hspace{1.1in} +\ \lambda_i^{LL} \lambda_j^{RL}x_jx_b\xi_g^{-}
      [x_i^2(g(x_i^2,x_j^2;1) - \frac{1}{4}f(x_i^2,x_j^2;1))
      (\bar{d_L}\gamma_\mu b_L)(\bar{d_R}\gamma_\mu b_R) \cr
  &&\hspace{2.2in} +\ (f(x_i^2,x_j^2;1) - x_i^2x_j^2g(x_i^2,x_j^2;1))
      (\bar{d_L}b_R)(\bar{d_R}b_L)]\biggr\} , \nonumber
\eeqr
where
\beq
\frac{G_F}{\sqrt{2}} \equiv \frac{g_L^2}{8M_W^2},\quad \xi_g^{\pm} \equiv e^{\pm\alpha}\xi_g ,
\quad \lambda_i^{AB} \equiv U_{id}^{A*}U_{ib}^{B} , \quad x_i \equiv \frac{m_i}{M_W} \ (i = u,c,t),
\eeq
and
\beqr
f(x_i,x_j;\zeta) = \frac{\ln (1/\zeta)}{(1-\zeta)(1-x_i\zeta)(1-x_j\zeta)}
 + \left(\frac{x_i^2\ln x_i}{(x_i-x_j)(1-x_i)(1-x_i\zeta)} + ( i \longrightarrow j )\right) ,\cr
g(x_i,x_j;\zeta) = \frac{\zeta\ln (1/\zeta)}{(1-\zeta)(1-x_i\zeta)(1-x_j\zeta)}
 + \left(\frac{x_i\ln x_i}{(x_i-x_j)(1-x_i)(1-x_i\zeta)} + ( i \longrightarrow j )\right) .
\eeqr
Although the form of the charged interactions in eqs.(\ref{Eff-HSM}-\ref{Eff-HLR}) is independent
of our particular choice of scalar representation, the Ward identities require that the box diagrams
contributing to $B^0\bar{B^0}$ mixing in the $LRM$ are not gauge invariant \cite{Chang84}.
In order to impose gauge invariance into our theory, we need to involve flavor-changing
neutral Higgs bosons, but it has been known that their contributions, even at the tree-level
as long as the mass of the flavor-changing Higgs boson is much heavier than
$M_{W^\prime}$,\footnote{The tree-level flavor-changing neutral Higgs contributions
with masses $M_H$ of order $M_{W^\prime}$ in the manifest or pseudo-manifest
left-right symmetric model were discussed in Ref. \cite{Moha83}.}
are suppressed by approximately a factor of $\zeta$ compared to the above gauge boson contributions
\cite{Hou}.  Therefore the above results, in the approximation of neglecting
external momenta and $d$-quark mass, provide the complete effective Hamiltonian contributing
to $B^0\bar{B^0}$ mixing.

At this stage, in order to analyze the obtained effective Hamiltonian quantitatively,
we need to consider specific forms of the right-handed quark mixing matrices $U^R$.
If the model has manifest or pseudo-manifest left-right symmetry, $W_R$ mass has a stringent
bound $M_{W_R}\geq 1.6$ TeV \cite{Beall}, and the $W_R$ boson contributions to $B^0\bar{B^0}$ mixing and
tree level b decay are very small.  But, in general, the form of $U^R$ is not necessarily to be
restricted to manifest or pseudo-manifest symmetric type, so the $W_R$ mass limit can be
lowered to approximately 300 GeV by taking the following forms of $U^R$ \cite{Olness84};
\beq
U^R_I = \left( \begin{array}{ccc} e^{i\omega} & \sim 0 & \sim 0 \\
                            \sim 0 & c_R e^{i\alpha_1} & s_R e^{i\alpha_2} \\
                       \sim 0 & -s_R e^{i\alpha_3} & c_R e^{i\alpha_4} \end{array} \right) ,\quad
U^R_{II} = \left( \begin{array}{ccc} \sim 0 & e^{i\omega} & \sim 0 \\
                            c_R e^{i\alpha_1} & \sim 0 & s_R e^{i\alpha_2} \\
                            -s_R e^{i\alpha_3} & \sim 0 & c_R e^{i\alpha_4} \end{array} \right) ,
\label{UR}
\eeq
where $c_R\ (s_R)\equiv \cos\theta_R\ (\sin\theta_R)$ $(0^\circ \leq \theta_R \leq 90^\circ )$.
Here the matrix elements indicated $\sim 0$ may be $\lesssim 10^{-2}$ and the unitarity
requires $\alpha_1+\alpha_4=\alpha_2+\alpha_3$. From the $b\rightarrow c$ semileptonic
decays of the $B$ mesons, we can get an approximate bound
$\xi_g\sin\theta_R \lesssim 0.013$ by assuming $|U^L_{cb}|\approx 0.04$ \cite{Voloshin97}.

The obtained effective Hamiltonians in eqs.(\ref{Eff-HSM}-\ref{Eff-HLR}) are then
further simplified using the Glashow-Iliopoulos-Maiani ($GIM$) cancellation
$\sum_{i=u,c,t}\lambda_i=0$ and neglecting the $u$-quark mass :
\beqr
H^{SM}_{eff} &=& \frac{G_F^2M_W^2}{4\pi^2}(\lambda_t^{LL})^2S(x^2_t)
                (\bar{d_L}\gamma_\mu b_L)^2 , \label{HSMeff}\\
H^{LR}_{eff} &=& \frac{G_F^2M_W^2}{2\pi^2}(\frac{g_R}{g_L})^2 \biggl\{
                  [\lambda_c^{LR} \lambda_t^{RL}x_cx_t\zeta A_1(x_t^2,\zeta)
                 + \lambda_t^{LR} \lambda_t^{RL}x_t^2\zeta A_2(x_t^2,\zeta)]
                   (\bar{d_L}b_R)(\bar{d_R}b_L) \cr
&&\hspace{0.2in} +\ \lambda_t^{LL} \lambda_t^{RL}x_b\xi_g^-
                    [x_t^3A_3(x_t^2)(\bar{d_L}\gamma_\mu b_L)(\bar{d_R}\gamma_\mu b_R)
                    + x_tA_4(x_t^2)(\bar{d_L}b_R)(\bar{d_R}b_L)]\biggr\} ,
\label{HLReff}
\eeqr
where
\beqr
S(x) &=& \frac{x(4 - 11x + x^2)}{4(1-x)^2} - \frac{3x^3\ln x}{2(1-x)^3} , \cr
A_1(x,\zeta) &=& \frac{(4-x)\ln x}{(1-x)(1-x\zeta)}
                 + \frac{(1-4\zeta)\ln \zeta}{(1-\zeta)(1-x\zeta)} , \cr
A_2(x,\zeta) &=& \frac{4-x}{(1-x)(1-x\zeta)} + \frac{(4-2x+x^2(1-3\zeta))\ln x}
                 {(1-x)^2(1-x\zeta)^2} + \frac{(1-4\zeta)\ln\zeta}{(1-\zeta)(1-x\zeta)^2} , \\
A_3(x) &=& \frac{7-x}{4(1-x)^2} + \frac{(2+x)\ln x}{2(1-x)^3} , \cr
A_4(x) &=& \frac{2x}{1-x} + \frac{x(1+x)\ln x}{(1-x)^2} . \nonumber
\eeqr
Note that $S(x)$ is the usual Inami-Lim function, $A_1(x,\zeta)$ is obtained by taking the limit
$x_c^2=0$, and $H_{eff}^{RR}$ is suppressed because it is proportional to $\zeta^2$.
Also, in the case of $U^R_I$, one can see that there is no significant contribution of
$H^{LR}_{eff}$ to $B^0\bar{B^0}$ mixing, so we will concentrate on the second type $U^R_{II}$
in this section.

The dispersive part of the $B^0\bar{B^0}$ mixing matrix element can then be written as
\beq \label{massmixing}
M_{12} = M_{12}^{SM} + M_{12}^{LR} = M_{12}^{SM}\biggl\{ 1 + \left(\frac{g_R}{g_L}\right)^2r_{LR} \biggr\} ,
\eeq
where
\beq
\left( \frac{g_R}{g_L} \right)^2r_{LR} \equiv \frac{M_{12}^{LR}}{M_{12}^{SM}}
  = \frac{<\bar{B^0}|H_{eff}^{LR}|B^0>}{<\bar{B^0}|H_{eff}^{SM}|B^0>} .
\eeq
For specific phenomenological estimates one needs the hadronic matrix elements of the operators
in eq.(\ref{HSMeff},\ref{HLReff}) in order to evaluate the mixing matrix element.
We use the following parametrization :
\beqr
<\bar{B^0}|(\bar{d_L}\gamma_\mu b_L)^2|B^0> &=& \frac{1}{3}B_1f_B^2m_B , \cr
<\bar{B^0}|(\bar{d_L}\gamma_\mu b_L)(\bar{d_R}\gamma_\mu b_R)|B^0>
  &=& -\frac{5}{12}B_2f_B^2m_B , \\
<\bar{B^0}|(\bar{d_L}b_R)(\bar{d_R}b_L)|B^0> &=& \frac{7}{24}B_3f_B^2m_B \nonumber,
\eeqr
where
\beq
<0|\bar{d_\beta}\gamma^\mu b_\alpha|B^0> = - <\bar{B^0}|\bar{d_\beta}\gamma^\mu b_\alpha|0>
 = - \frac{if_Bp_B^\mu}{\sqrt{2m_B}}\frac{\delta_{\alpha\beta}}{3} ,
\eeq
and where $f_B$ is the $B$ meson decay constant and $B_i\ (i=1,2,3)$ are the bag factors.
In the vacuum-insertion method \cite{Gaillard}, $B_i=1$ in the limit $m_b\simeq m_B$.
We will use $f_BB_i^{1/2}=(210\pm 40)$ MeV for our numerical estimates \cite{Bernard}.
Using the standard values of the quark masses and
$|U^L_{cd}|\approx 0.222$, one can express $r_{LR}$
in terms of the mixing angle and phases in (\ref{UR}) as
\beqr \label{rLR}
r_{LR} &\approx& \textit{l} \biggl\{ 18.1 \textit{l} \biggl(
 \frac{1 - \zeta - (3.49 - 14.0\zeta)\ln(1/\zeta)) }{ 1 - 5.68\zeta }\biggr)
 \zeta s_R^2e^{i\delta_1} \cr
 &-& 739 \biggl( \frac{ 1 - 5.04\zeta - (0.483 - 1.93\zeta )\ln(1/\zeta) }
         {1 - 10.4\zeta + 31.3\zeta^2}\biggr) \zeta s_Rc_Re^{i\delta_2} \
         - \ 7.68\xi_g s_Re^{i\delta_3} \biggr\} ,
\eeqr
where $\textit{l} = 0.009/|U^L_{td}|$, $\delta_1 = -2\beta + \alpha_2 - \alpha_3$,
$\delta_2 = -\beta - \alpha_3 + \alpha_4$, $\delta_3 = -\beta - \alpha_3$ and the mixing phase
$\alpha$ was absorbed in $\alpha_i$ by redefining $\alpha_i + \alpha \rightarrow \alpha_i$.

\begin{figure}[!hbt]
  \vskip3mm
  \centerline{\resizebox{8cm}{!}{%
  \includegraphics{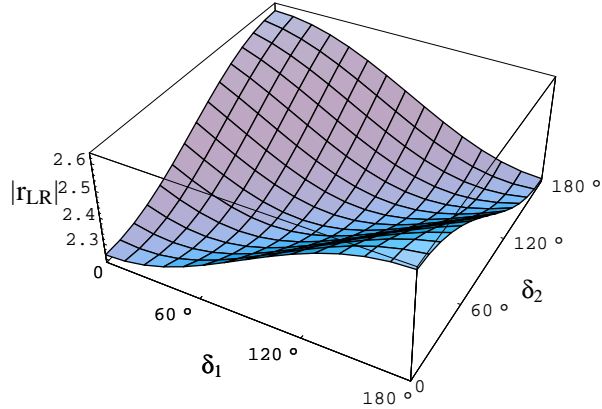}}}
  \caption{Behavior of the ratio $|r_{LR}|$ as $\delta_{1,2}$ are varied.}
  \label{fig:rLRphase}
\end{figure}
\begin{figure}[!hbt]
  \vskip3mm
  \centerline{\resizebox{8cm}{!}{%
  \includegraphics{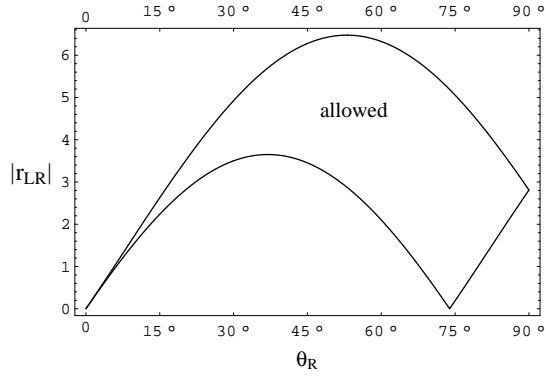}}}
  \caption{Allowed region for $|r_{LR}|$ and $\theta_R$.}
  \label{fig:rLRangle}
\end{figure}

\begin{figure}[!hbt]
  \vskip3mm
  \centerline{\resizebox{13cm}{!}{%
  \includegraphics{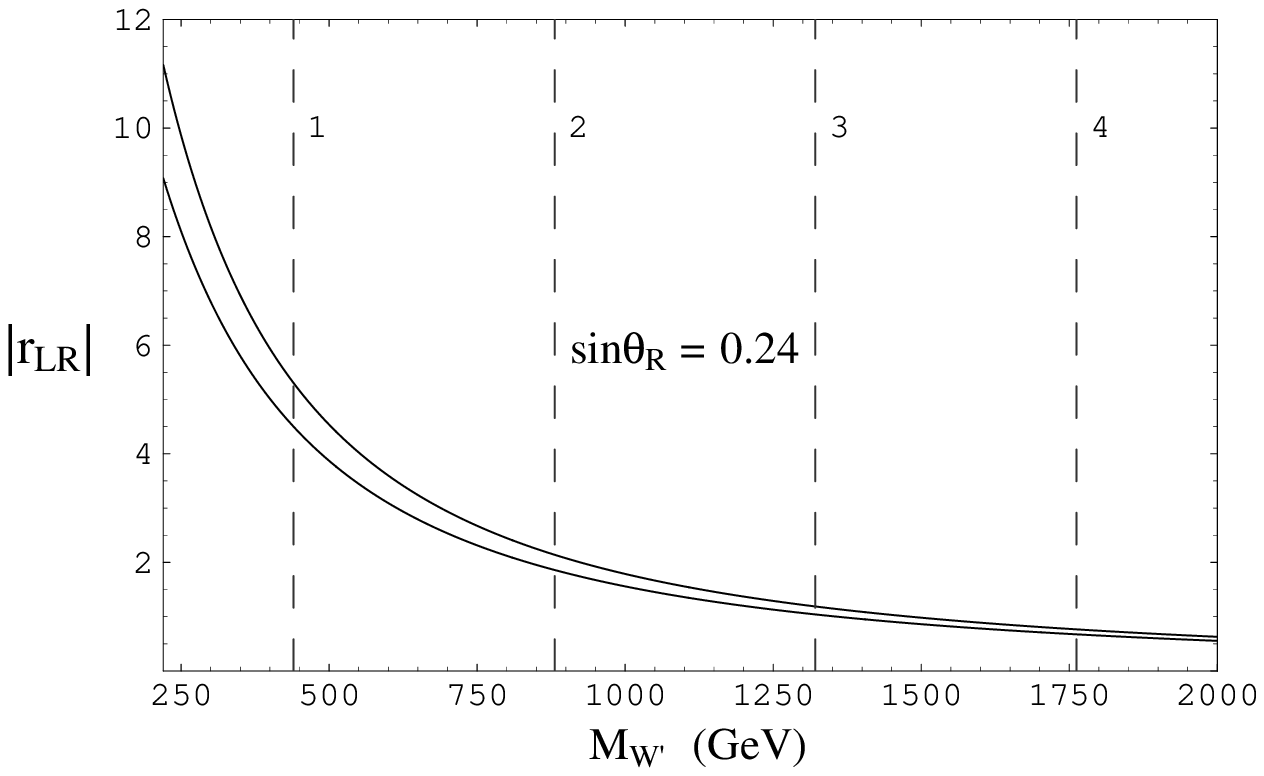}\hfill
  \includegraphics{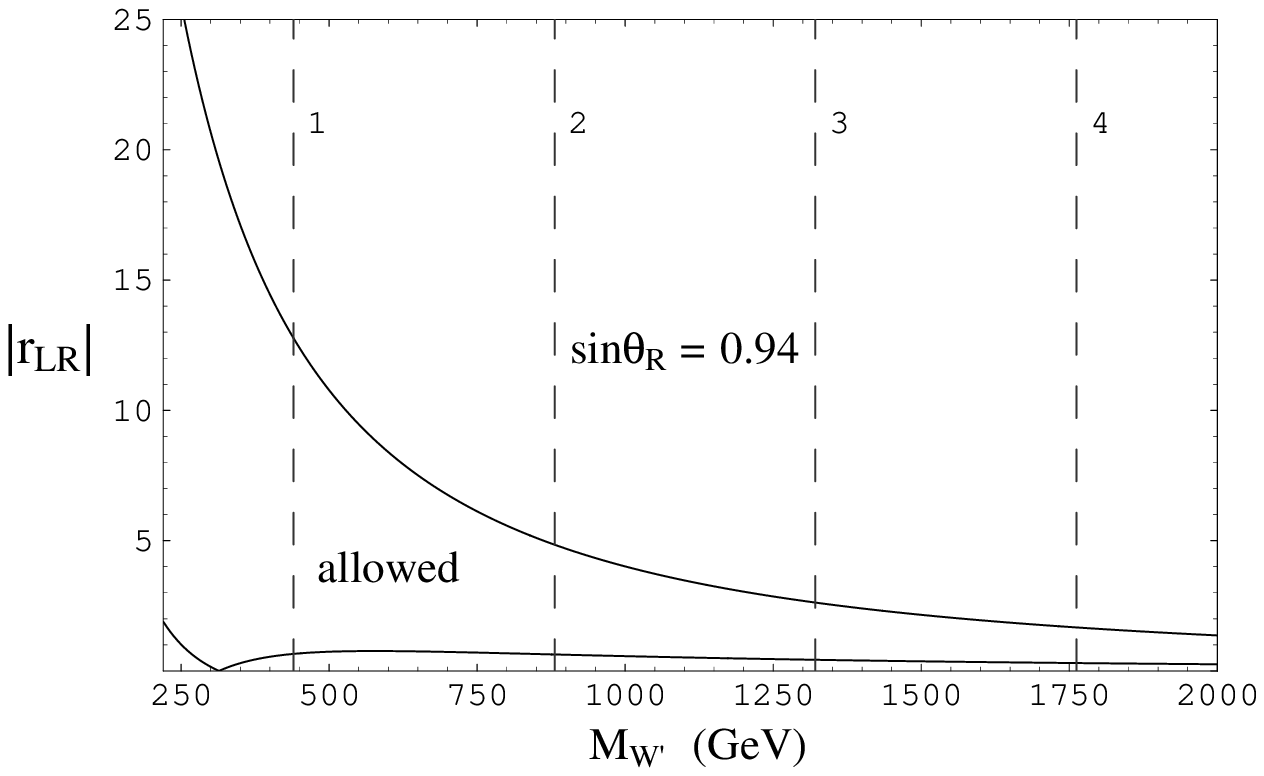}}}
  \caption{Allowed regions for $|r_{LR}|$ and $M_{W^\prime}$ for $g_R/g_L \geq 0.5$.
   The dotted lines correspond to the lower bounds on $M_{W^\prime}$
   in eq.(\ref{MWRbound}) for the ratio $g_R/g_L$ = 1,2,3 and 4 respectively. }
  \label{fig:rLRM}
\end{figure}

\begin{figure}[!hbt]
  \vskip3mm
  \centerline{\resizebox{13cm}{!}{%
  \includegraphics{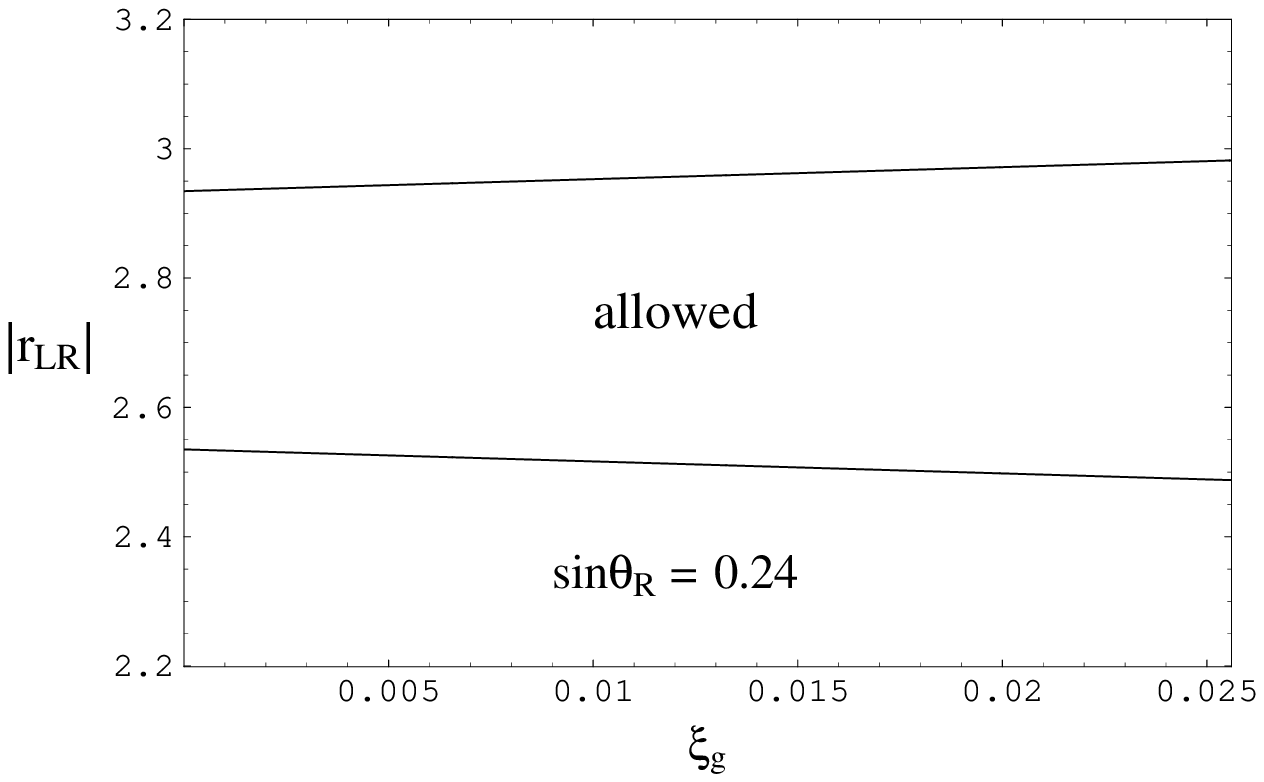}\hfill
  \includegraphics{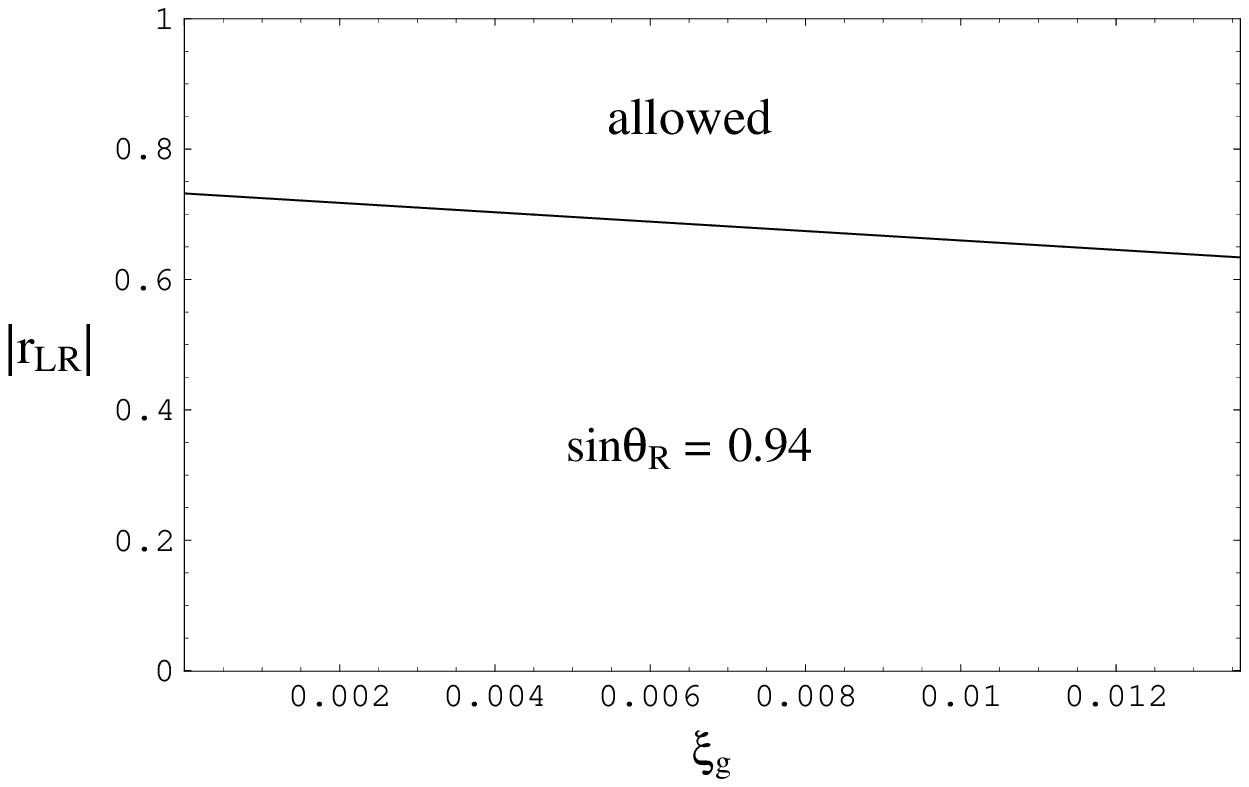}}}
  \caption{Allowed regions for $|r_{LR}|$ and $\xi_g$ for $M_{W^\prime} =$ 700 GeV.}
  \label{fig:rLRX}
\end{figure}

  Now we investigate numerically the behavior of the ratio $|r_{LR}|$,
which is the deviation of $M_{12}$ from the $SM$,
under variation of $M_{W^\prime}$, $\xi_g$, $\theta_R$ and the phases
in $U^R$, assuming $\textit{l} = 1$.  Although we use the average value of $|U_{td}|$
which might be different from the actual value of $|U^L_{td}|$, it should not affect
the order of magnitude in our estimates.
First, in order to see the dependence of $|r_{LR}|$ on the phases,
we fix $M_{W^\prime} =$ 800 GeV, $\xi_g = 0.005$, $\theta_R = 15^\circ$,
and set $\delta_3 = \pi$ because its effect is relatively much smaller than that of
$\delta_1$ and $\delta_2$.  The plot is shown in Fig.\ref{fig:rLRphase}.
From eq.(\ref{rLR}) and Fig.\ref{fig:rLRphase}, one can see that $|r_{LR}|$ becomes maximal
when $\delta_{1,3} = \pi$ and $\delta_2 = 0$ ,
and minimal when $\delta_{1,2,3} = \pi$ if $\theta_R \lesssim 70^\circ$
(or $\delta_{1,2} = \pi$ and $\delta_3 = 0$ if $\theta_R \gtrsim 70^\circ$ ).
This behavior also holds for other values of $M_{W^\prime}$ and $\xi_g$.
 Since $|r_{LR}|$ is the continuously varying function of the phases, we can probe the allowed
region for $|r_{LR}|$ with respect to the parameters $M_{W^\prime}$, $\xi_g$ and $\theta_R$.
Next, we fix $M_{W^\prime} = $ 800 GeV, $\xi_g = 0.005$, and evaluate $|r_{LR}|$ by varying $\theta_R$.
Note that $|r_{LR}|$ can approach zero at a non-zero $\theta_R$ near $73^\circ$ as shown
in Fig.\ref{fig:rLRangle}.
Otherwise, it is larger than 1, which means that generally it is possible to have
$|M_{12}^{LR}| \gg |M_{12}^{SM}|$.
In Fig.\ref{fig:rLRM}, we consider the behavior of $|r_{LR}|$ for $g_R/g_L \geq 0.5$,
$\xi_g = 0.0004$ and $\theta_R = 14^\circ, 70^\circ$ as $M_{W^\prime}$ is varied.
The behavior of $|r_{LR}|$ exhibits a substantial dependence on $M_{W^\prime}$,
and $|r_{LR}|$ can be larger than 1 even for $M_{W^\prime} \sim 2$ TeV.
Moreover, it can be noticed that $|r_{LR}|$ falls near $M_{W^\prime} \sim 300$ GeV at certain
angles and phases in the mixing matrices. This reflects a possibility of relatively light
masses of $W^\prime$ compared to the previously known bound.
We will return to this point in Sec.4.  The dependence of $|r_{LR}|$ on $\xi_g$ satisfying
$\xi_g\sin\theta_R \lesssim 0.013$
at fixed $M_{W^\prime} = 700$ GeV and $\theta_R = 14^\circ, 70^\circ$
is shown in Fig.\ref{fig:rLRX}.  As one can see, $|r_{LR}|$ can be enhanced up to 10\%
of the $SM$ contribution for the given inputs.  Although its effect is smaller than
that of other parameters, it is not negligible and can be dominant in $|r_{LR}|$
if the first two $\zeta$ dependent terms in eq.(\ref{rLR}) cancel each other.


As we mentioned previously, the average value of $|U_{td}|$ might be different from the actual
value of $|U^L_{td}|$, and there is also ambiguity from errors in $f_BB_i^{1/2}$.
Therefore the mass mixing $\Delta M_B^{SM}$ can be either much larger or smaller
than $\Delta M_B^{exp}$.  However, if we assume that
$0.5 \lesssim |\Delta M_B^{SM}/\Delta M_B^{exp}| \lesssim 2$,  we can get specific bounds on
the mass $M_{W^\prime}$ and the angle $\theta_R$ using the experimental value
$\Delta M^{exp}_B \simeq 0.472\times 10^{12}s^{-1}$.
We will estimate the lowest possible bound on $M_{W^\prime}$ with respect to $\theta_R$
in their parameter space with the numerical consideration of $\sin 2\beta$
in the next section.

\section{CP asymmetry in $B^0$ decay}
The $CP$ angle $\beta$ in $CKM$ matrix can be measured in $B \rightarrow J/\psi K_S$ decays.
In $B$ decays into a final $CP$ eigenstate $J/\psi K_S$, $\beta$ is related
to a parametrization invariant quantity $\lambda$ as follows \cite{Review}:
\beq
\sin 2\beta_{eff} =  \textrm{Im} \lambda (B^0 \rightarrow J/\psi K_S ) ,
\eeq
where
\beq
\lambda \equiv - \left( \frac{q}{p}\right)_B \frac{A(\bar{B^0}
  \rightarrow J/\psi K_S )}{A(B^0 \rightarrow J/\psi K_S )} , \qquad \left(\frac{q}{p}\right)_B
  \simeq \frac{M^*_{12}}{|M_{12}|} .
\eeq
The minus sign in the above expression is coming from the fact that $J/\psi K_S$ is CP-odd.
As mentioned earlier, $\beta_{eff}=\beta$ in the $SM$.

In the $LRM$, the two types of $U^R$ give us two distinct results.  In the case of $U^R_I$,
the $W^\prime$ contribution to the mixing parameter $(q/p)_B$ is negligible so that
$(q/p)_B \simeq (q/p)_{SM} = e^{-2i\beta}$. Then the $CP$ angle $\beta_{eff}$ can be expressed
by
\beqr \label{sin2b1}
\sin 2\beta_{eff}^I &\simeq & - \textrm{Im}\left( e^{-2i\beta}
 \frac{U_{cs}^{L\ast}U_{cb}^{L} + (g_R/g_L)^2 (-2U_{cs}^{L\ast}U_{cb}^{R}\xi_g^+
                                               + U_{cs}^{R\ast}U_{cb}^{R}\zeta )}
      {U_{cs}^{L}U_{cb}^{L\ast} + (g_R/g_L)^2 (-2U_{cs}^{L}U_{cb}^{R\ast}\xi_g^-
                                               + U_{cs}^{R}U_{cb}^{R\ast}\zeta )}\right) \cr
&\simeq & - \textrm{Im}\left( e^{-2i\beta}
 \frac{1 + 25(g_R/g_L)^2 (-2s_R\xi_ge^{i\alpha_2} + c_Rs_R\zeta e^{i(\alpha_2-\alpha_1)})}
 {1 + 25(g_R/g_L)^2 (-2s_R\xi_ge^{-i\alpha_2} + c_Rs_R\zeta e^{-i(\alpha_2-\alpha_1)})}\right) ,
\eeqr
where the mixing angle $\alpha$ is absorbed in $\alpha_i$ again, and we ignored
the $K\bar{K}$ mixing and assumed that
\beqr
<J/\psi K_s|\bar{c}_L\gamma_\mu s_L \bar{b}_L\gamma^\mu c_L|B^0>
 &\simeq& <J/\psi K_s|\bar{c}_R\gamma_\mu s_R \bar{b}_R\gamma^\mu c_R|B^0> \cr
 &\simeq& - \frac{1}{2}<J/\psi K_s|\bar{c}_L\gamma_\mu s_L \bar{b}_R\gamma^\mu c_R|B^0> .
\eeqr
As one can easily see in eq.(\ref{sin2b1}), $\sin 2\beta_{eff}$ = $\sin 2\beta$
if $\alpha_{1,2} = 0$ or $\pi$.

\begin{figure}[!hbt]
  \vskip3mm
  \centerline{\resizebox{8cm}{!}{%
  \includegraphics{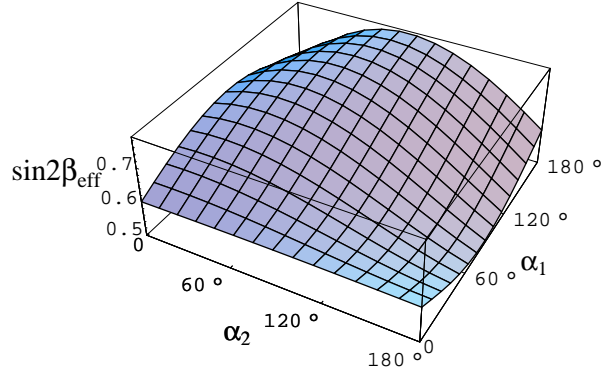}}}
  \caption{Behavior of $\sin 2\beta_{eff}$ as $\alpha_{1,2}$ are varied.}
  \label{fig:asym11}
\end{figure}

\begin{figure}[!hbt]
  \vskip3mm
  \centerline{\resizebox{7cm}{!}{%
  \includegraphics{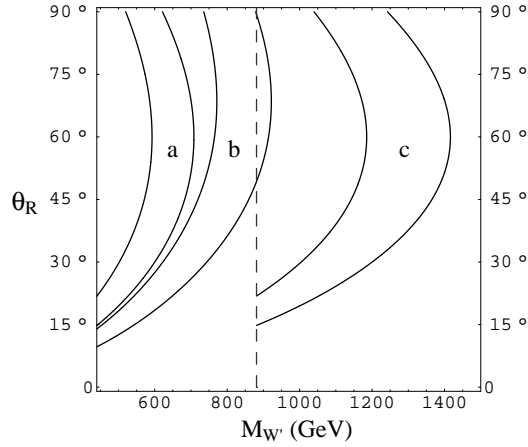}}}
  \caption{Contour plots corresponding to $\sin 2\beta_{eff} = 0.99$
   for $\sin 2\beta = 0.60$, (a) $\xi_g = \zeta/2$ and $g_R = g_L$, (b) $\xi_g = \zeta$
   and $g_R = g_L$, and (c) $\xi_g = \zeta/2$ and $g_R = 2g_L$.
  The dotted line corresponds to the lower bound on $M_{W^\prime}$ for $g_R = 2g_L$.}
  \label{fig:asym12}
\end{figure}

For illustration of the possible effect of the new interaction on the effective value
of $\sin 2\beta$ : $\sin 2\beta_{eff}$, we assume that the $SM$ contribution produces
$\sin 2\beta$ = 0.60, and show the region of parameters, where the effective value is
shifted to $\sin 2\beta_{eff} \sim 1$.
We first plot $\sin 2\beta_{eff}$ in Fig.\ref{fig:asym11} for the typical values
$M_{W^\prime} = 800$ GeV, $\xi_g = 0.005$, $\theta_R = 15^\circ$ and $g_R = g_L$
as $\alpha_{1,2}$ are varied.  In the figure, $\sin 2\beta_{eff}$ has a maximum variation from
$\sin 2\beta$ near $\alpha_2 \approx \pi/2$ and $\alpha_1 = \pi$,
and this behavior holds for other values of $M_{W^\prime}$, $\xi_g$ and $\theta_R$.
Next, we plot the contour corresponding to $\sin 2\beta_{eff} = 0.99$ satisfying
$\xi_g\sin\theta_R \lesssim 0.013$ in the parameter space of $M_{W^\prime}$
and $\theta_R$ for $\alpha_2 = \pi/2$, $\alpha_1 = \pi$, $\xi_g = \zeta/2, \zeta$
and $g_R/g_L = 1, 2$ in Fig.\ref{fig:asym12}.\footnote{We did the same analysis
for $g_R/g_L = 0.5$ but there was no allowed region.}
As one can see, the upper bound of $M_{W^\prime}$ goes down with decreasing
$\xi_g$ and $g_R/g_L$.  Therefore, under the given assumption, $g_R \ll g_L$ is disfavored,
and so is $M_{W^\prime} \gg $ 1 TeV unless $g_R \gg g_L$.

\begin{figure}
 \vskip3mm
 \begin{minipage}[!hbt]{16cm}
  \centerline{\resizebox{11cm}{!}{%
  \includegraphics{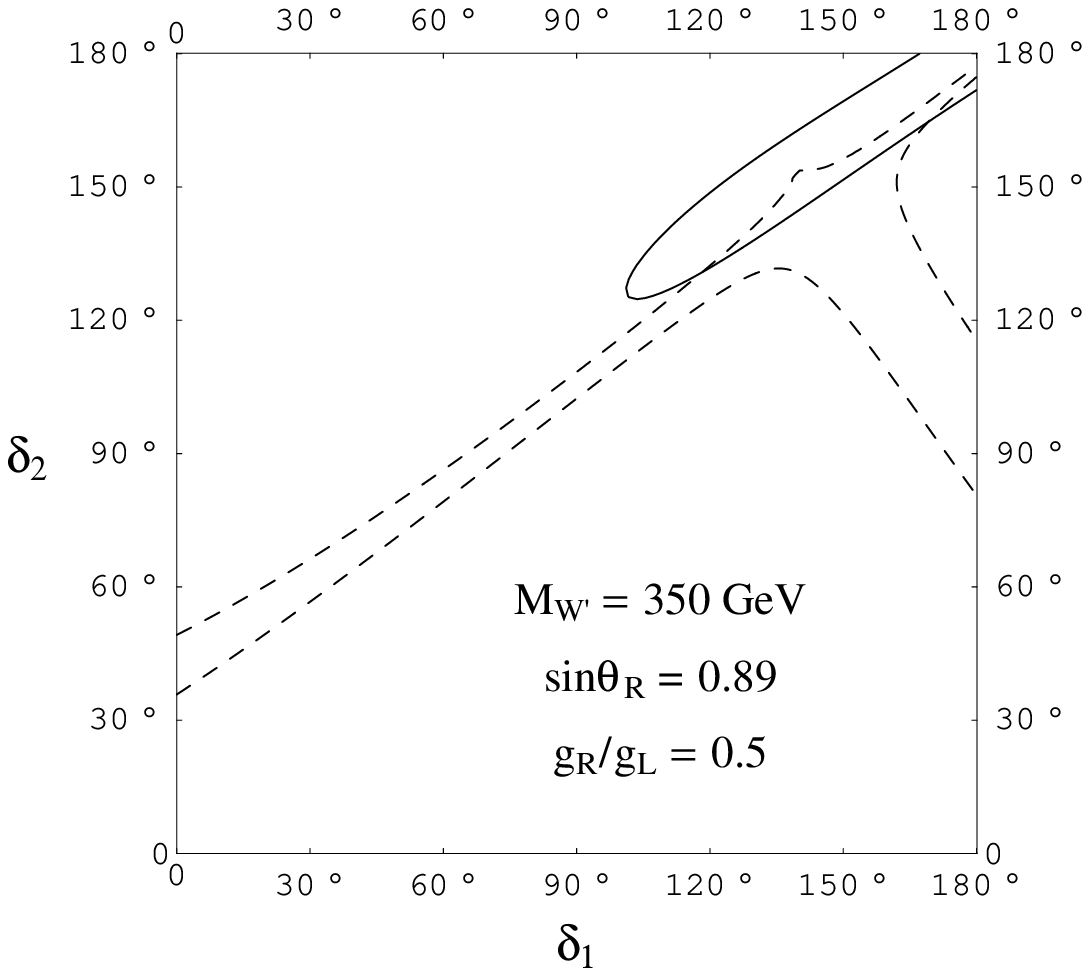} \qquad
  \includegraphics{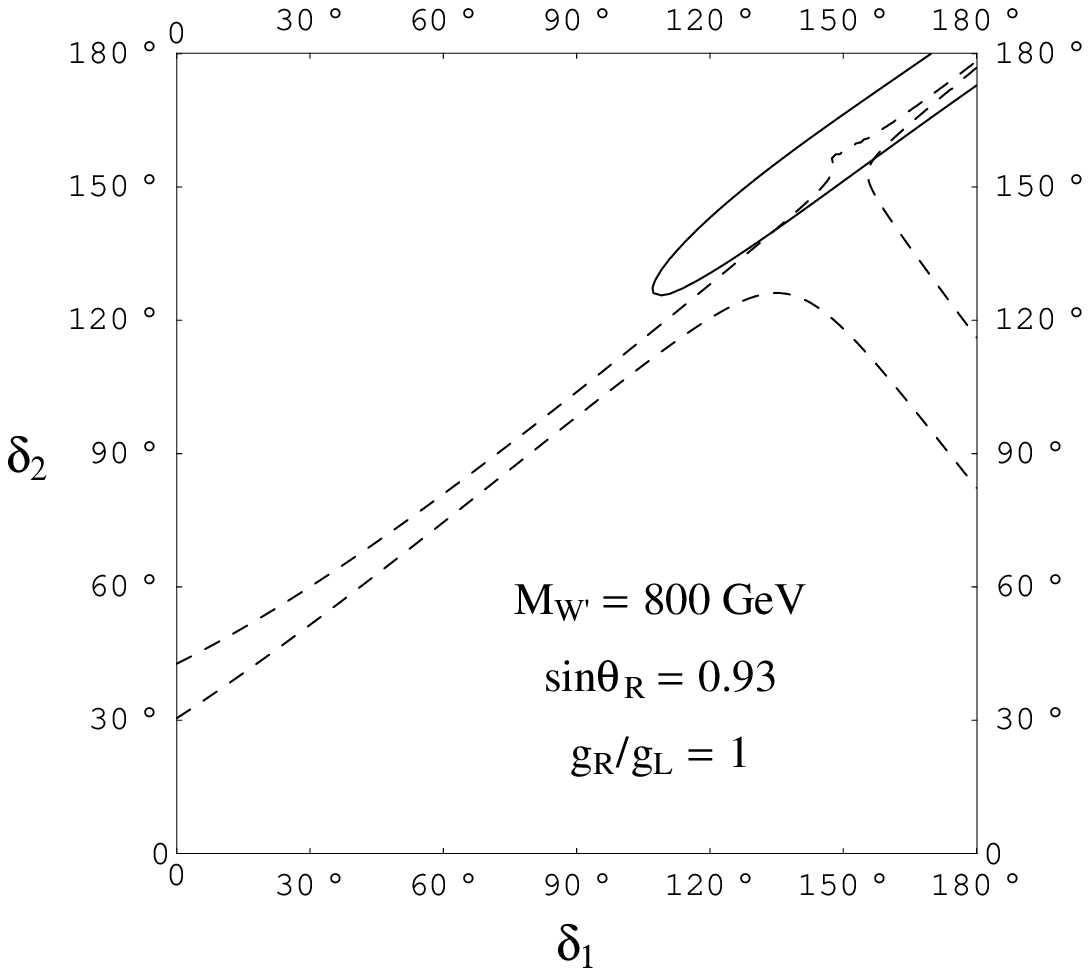}}}
 \end{minipage} \\
 \begin{minipage}[!hbt]{16cm}
  \centerline{\resizebox{11cm}{!}{%
  \includegraphics{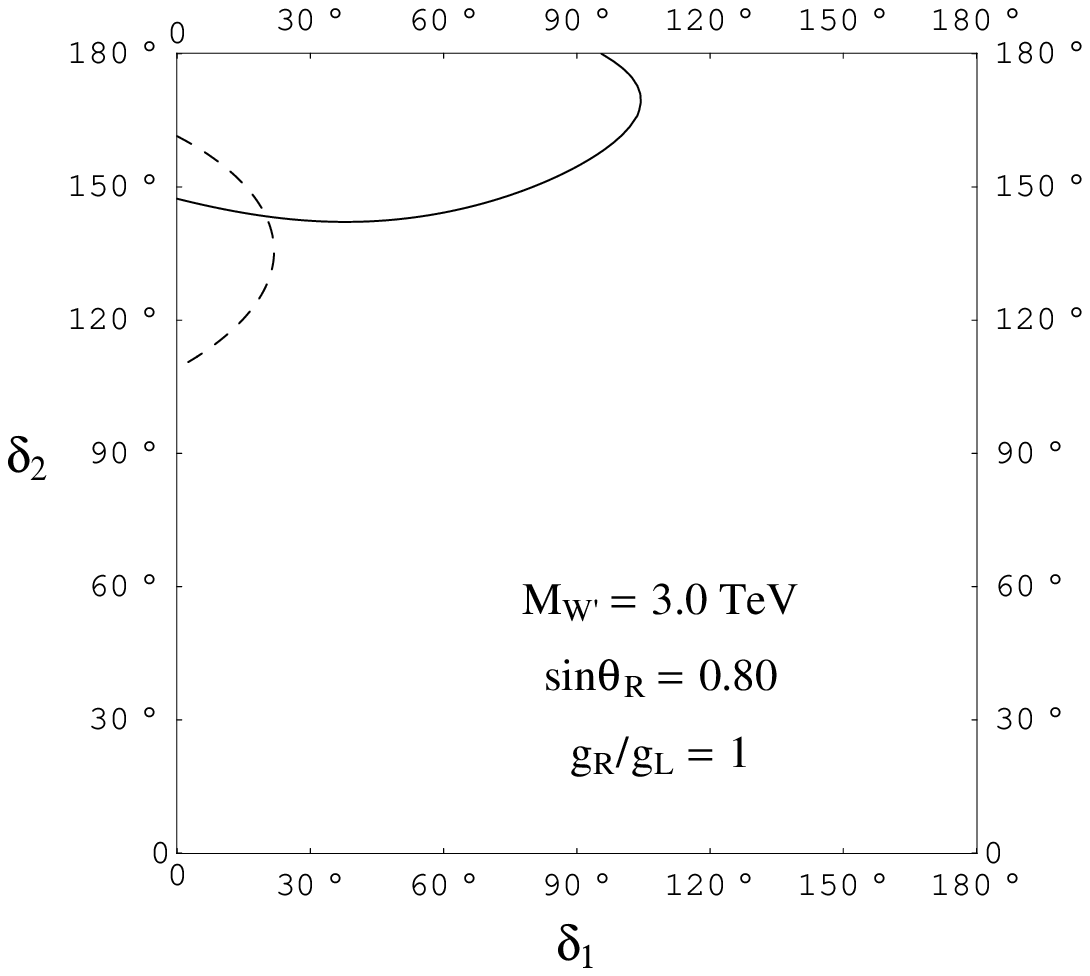} \qquad
  \includegraphics{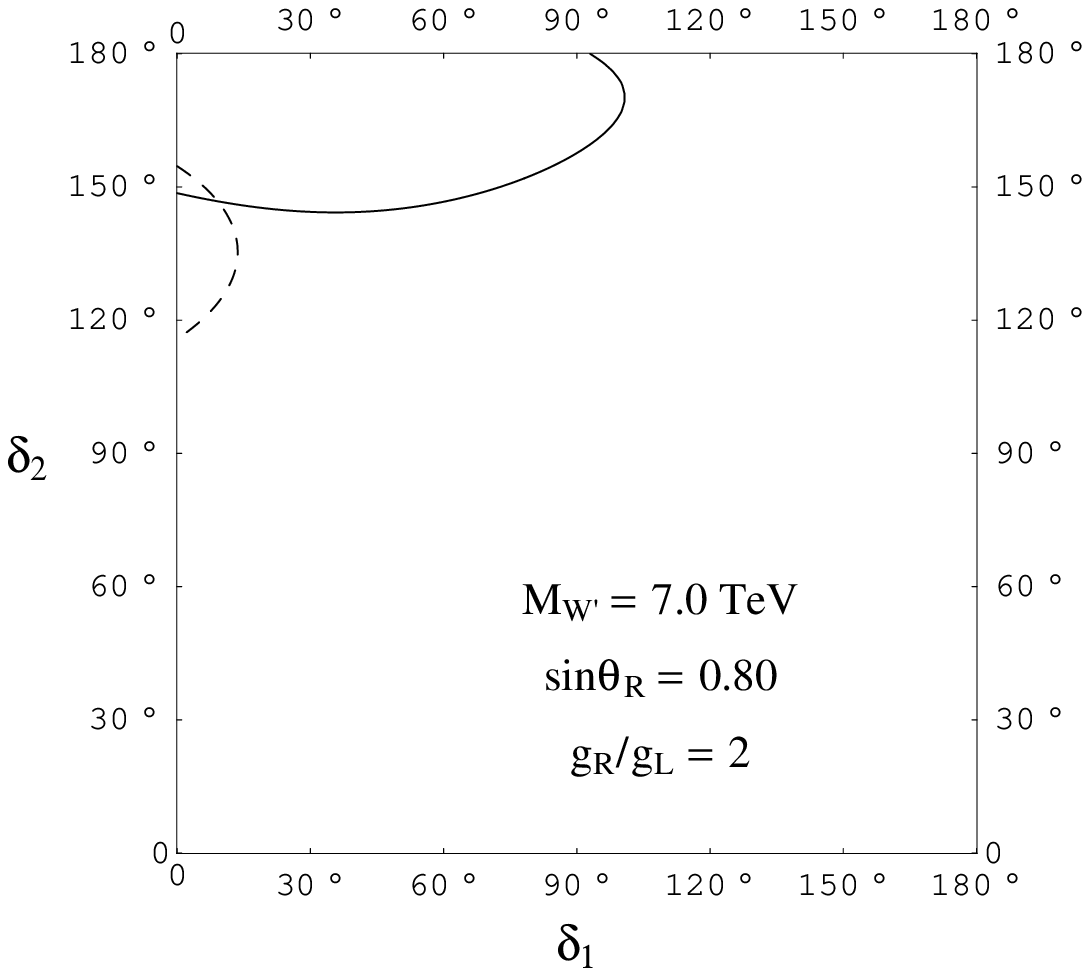}}}
 \end{minipage}
  \caption{Contour plots of $\Delta M_B$ (dashed) for $|U^L_{td}| = 0.009$ and
  those corresponding to $\sin 2\beta_{eff} = 0.99$ (dotted) for $\sin 2\beta = 0.60$.}
  \label{fig:asym2}
\end{figure}

In the case of $U^R_{II}$, one has $U^R_{cs}\sim 0$ so that the $\zeta\ (M_{W^\prime})$ dependent term
in eq.(\ref{sin2b1}) is very small. However the ratio $(q/p)_B$ depends on $M_{W^\prime}$.
Thus the $W^\prime$ contribution enters in a somewhat different way.
\beq \label{sin2b2}
\sin 2\beta_{eff}^{II}\ \simeq \ - \textrm{Im}\left( e^{-2i\beta}
 \frac{(1 + (g_R/g_L)^2 r_{LR}^\ast )}{|1 + (g_R/g_L)^2 r_{LR} |}
 \frac{(1 - 50(g_R/g_L)^2 s_R\xi_ge^{i\alpha_2} )}
 {(1 - 50(g_R/g_L)^2 s_R\xi_ge^{-i\alpha_2} )}\right) .
\eeq

Unlike in the previous case, we need to consider here the mass mixing of $B^0\bar{B^0}$
in order to analyze $\sin 2\beta_{eff}$ numerically.
Assuming that $|U^L_{td}| = 0.009$ and $\sin 2\beta = 0.60$, we plot
the contours corresponding to $\sin 2\beta_{eff} = 0.99$ and $\Delta M_B^{LR} = \Delta M_B^{exp}$
in the parameter space of $\delta_{1,2}$ for $\delta_3 = \pi$, $\xi_g = \zeta/4$ and
$g_R/g_L = 0.5,1,2$ by varying $\theta_R$ and $M_{W^\prime}$ in Fig.\ref{fig:asym2}.
Because of the non-triviality of the behavior of $\sin 2\beta_{eff}$ on $\delta_i$,
we repeated this analysis until the two contours have overlapped by varying $M_{W^\prime}$
from 350 GeV to 8 TeV, and found that it appeared
 where 350 GeV $\lesssim M_{W^\prime} \lesssim$ 1.3 TeV if $g_R/g_L = 0.5$,
440 GeV $\lesssim M_{W^\prime} \lesssim$ 3.1 TeV if $g_R/g_L = 1$ and
880 GeV $\lesssim M_{W^\prime} \lesssim$ 7.1 TeV if $g_R/g_L = 2$.
Even though the existence of a heavy $W^\prime$
with the mass $M_{W^\prime} > $ 7 TeV may be allowed by the numerical analysis of $\Delta M_B^{LR}$,
it is excluded by that of $\sin 2\beta_{eff}$ under the given assumption.
For the different values of $\xi_g$, we also have similar results.

\section{Conclusion}
In the $LRM$, if one does not impose the manifest or pseudo-manifest left-right symmetry,
the $W^\prime$ contributions to $B^0\bar{B^0}$ mixing and $CP$ asymmetry in $B^0$ decays
are highly dependent upon the phases in the mass mixing matrix $U^{L,R}$.
For certain phases, the contribution of $W^\prime$ with a heavy mass about a few TeV
to $B^0\bar{B^0}$ mixing can be sizeable.  On the other hand, there is also a possibility of
the existence of $W^\prime$ with a light mass about a few hundred GeV, whose contribution
can be either very large or small, and so the contribution of the mixing angle $\xi$ is
not negligible.
Since the existence of a light $W^\prime$ requires a small $g_R$, $g_R \lesssim g_L$, one can
see from eq.(\ref{massmixing}) and Fig.\ref{fig:rLRM} that its contribution is limited.
Therefore even assuming that $\Delta M^{LR}_B \lesssim \Delta M^{SM}_B$, we find that
there is a possibility of a light $W^\prime$ with a mass $M_{W^\prime} \sim $ 300 GeV.

Also this possibility arises from the numerical analysis of the $CP$ asymmetry in $B^0$ decay.
Since $U_{td}^L$ is not known with sufficient accuracy, estimates of the pure right-handed
current contributions to $\Delta M_B$ and $\sin 2\beta$ are somewhat uncertain.
But, for the certain values of the parameter sets, one can see from Fig.\ref{fig:asym12}
and Fig.\ref{fig:asym2} that the $CP$ asymmetry parameter $\sin 2\beta$ can be as large as
almost 1, and the mass of $W^\prime$ can be as small as about 350 GeV.
Therefore, the existence of the light $W^\prime$ can be tested
once future experiments confirm the value of $\sin 2\beta$ and $|U_{td}^L|$.

\section{Acknowledgements}
The author would like to thank M. B. Voloshin for helpful comments.

\newpage

\end{document}